\documentclass[preprint, english]{nrc1}

\usepackage[english]{babel}

\usepackage{epsfig,multicol} 
\usepackage{amsmath}
\usepackage{amsfonts}
\usepackage{textcomp}

\journal{Can. J. Phys.}

\def\refpos#1 #2 #3{\global\xrefpos=#1 \global\yrefpos=#2
                         \rlap{$\smash{#3}$}}
\def\put #1 #2 #3{\xput=#1 \yput=#2
                  \advance\xput by -\xrefpos
                  \advance\yput by -\yrefpos
                  \rlap{\kern\the\xput truebp
                        \vbox to 0pt{\vss\hbox{$\displaystyle #3$}
                        \kern\the\yput truebp}}}
\def\beginlabels\refpos#1\endlabels{\hbox{$\refpos#1$}}

\renewcommand{\theequation}{\arabic{section}.\arabic{equation}}

\newcommand{\be}{\begin{equation}}
\newcommand{\ee}{\end{equation}}
\newcommand{\bea}{\begin{eqnarray}}
\newcommand{\eea}{\end{eqnarray}}

\begin{document}

\title{Holography and Fermions at a Finite Chemical Potential}

\author{Moshe Rozali}

\address{Department of Physics and Astronomy, University of British Columbia\\
Vancouver, BC, V6T 1Z1, Canada\\
\email{rozali@phas.ubc.ca}
}

\author{Lionel Brits}
\address{Department of Physics and Astronomy, University of British Columbia\\
Vancouver, BC, V6T 1Z1, Canada\\
\email{lionelbrits@phas.ubc.ca}
}
\shortauthor{M. Rozali, L. Brits}



\maketitle

\begin{abstract}
We review the Sakai-Sugimoto model of holographic QCD at zero temperature and finite chemical potential, comparing the results to those expected at large-$N_c$ QCD, and those in a closely related  holographic model. We find that as the baryon chemical potential is increased above a critical value, there is a phase transition to a nuclear matter phase, the details of which depend on the model. We argue that the nuclear matter phase is necessarily inhomogeneous to arbitrarily high density, which suggests an explanation of the ``chiral density wave'' instability of the quark Fermi surface in large-$N_c$ QCD. Some details of the instanton distribution in the holographic dual are reminiscent of a Fermi surface. This short manuscript summarizes a talk given by M.R. at ``Theory Canada 4'' conference, and is based largely (but not entirely) on the results of \cite{Rozalietal2008}.
\end{abstract}


\section{Introduction}

The phase diagram of QCD as a function of temperature and baryon chemical potential displays a rich variety of phases. A better understanding of the details of the phase diagram at intermediate chemical potentials would have valuable applications, for example in understanding the physics of neutron-star interiors. Currently our knowledge of that regime of the phase diagram is based exclusively on extrapolations and semi-empirical toy models. However, with the advent of  gauge-gravity dualities, we have a new tool for studying the properties of certain strongly coupled gauge theories. It is interesting to study these QCD-like theories in regimes for which neither analytic or numerical studies are currently possible in real QCD. In the present paper, we will focus then on equilibrium properties of those holographic models, at zero temperature and finite baryon chemical potential.

Additionally, the study of the phase diagram of QCD involves concepts common in many condensed matter systems, such as Fermi surfaces and their instabilities, superconductivity, and inhomogeneous phases. Therefore, the holographic study of the phase diagram is a natural first step towards a better understanding of holographic duals of these interesting phenomena in the context of strongly interacting condensed matter systems.

In order to get behavior similar to real QCD in the regime discussed here, it seems essential to study a theory with baryon charge carried exclusively by fermionic fields, otherwise the physics at finite chemical potential is likely to involve Bose condensation rather than the formation of a Fermi surface. Such a model was constructed a few years ago by Sakai and Sugimoto~\cite{Sakai:2004cn}, and will be focused on here. In addition to reviewing the results in \cite{Rozalietal2008}, we also provide a point of comparison by presenting results in parallel for a similar model, defined below, which has flavoured boson-fermion degeneracy and is therefore expected to have different qualitative behavior. We present the qualitative results of this comparison below.

\section{Overview of Models}
\subsection{The Adjoint Sector}
The basic setup for our models begins with the low-energy decoupling limit of $N_c$ D4-branes wrapped on a circle of length $2 \pi R$, with anti-periodic boundary conditions for the fermions. Apart from $N_c$, this theory has a single dimensionless parameter
$\lambda $,
the four-dimensional 't Hooft coupling at the Kaluza-Klein scale. As pointed out by Witten~\cite{Witten:1998zw}, at small $\lambda$ the dynamics should be that of pure Yang-Mills theory. Unfortunately, in this limit, the dual gravity background is highly curved, and so we are not in a position to study it perturbatively. For large $\lambda$, on the other hand, the gravity background is weakly curved, and so studying the theory is possible by using classical gravity calculations. This is a general phenomenon in the gauge-gravity duality; the gravity dual necessarily deviates from an asymptotically free field theory in the ultraviolet, and can be thought of as strong (bare) coupling expansion.

In the large-$N_c$ limit the gravity dual is the near-horizon geometry of the D4-branes. The metric is~\cite{Witten:1998zw,Kruczenski2004}
\begin{subequations}
\begin{align}
\label{D4metric}
ds^2 &= \left(\frac{U}{R_4} \right)^{\tfrac{3}{2}} \left(\eta_{\mu\nu} dx^\mu dx^\nu + f(U) dx_4^2\right) + \left(\frac{R_4}{U} \right)^{\tfrac{3}{2}}\left( \frac{dU^2}{f(U)} + U^2 d\Omega_4^2\right),\\
 f(U) &= 1-\left( \frac{U_0}{U}\right)^3.
\end{align}
\end{subequations}
where $d\Omega_4$ is the volume-element of a unit 4-sphere. The $x_4$ direction, corresponding to the Kaluza-Klein direction in the field theory, is taken to be asymptotically periodic, with coordinate periodicity $2 \pi R$. It combines with the radial coordinate $u$ to form a cigar-like geometry, with the topologically of a disk.

The study of the pure Yang-Mills theory holographically is similar to AdS/CFT; for example, one has the following qualitative features:
\begin{itemize}
\item Harmonics of closed string fields represent the glueball spectrum, while those of open strings, introduced below, are related to the meson spectrum.
    \item There is a deconfinement phase transition, related to the fact that there are two Euclidean geometries contributing to the finite temperature partition function. In fact, those two geometries are related by a ``flop'' -- an exchange of the compact direction $x_4$ with the Euclidean time.
\end{itemize}
\subsection{Adding Flavours}

Now that we have defined the adjoint sector of the theory, we would like to add $N_f$ flavours of fundamental quarks. By keeping $N_f$ fixed in the large-$N_c$ limit, we work in the so-called ``quenched approximation'', whereby the addition of matter (in the form of probe branes) does not modify the background geometry of the gravity dual.

The Sakai-Sugimoto construction is motivated by the observation that the light open string modes which live at the \((3+1)\)-dimensional intersection of a particular configuration of D4- and D8-branes give rise to chiral fermion fields on the intersection without accompanying bosons. Thus, to the original D4-branes, which we can take to lie in the 01234 directions, one adds a stack of $N_f$ D8-branes forming a line in the cigar spanned by $(U,x_4)$. The configuration is represented by the following table:
\begin{center}
\begin{tabular}{c|ccccccc}
     & $x^0$ & $x^1$ & $x^2$ & $x^3$ & $(x^4)$ & $U$ & $\Omega_4$ \\
\hline
 D4: & $-$ & $-$ & $-$ & $-$ & $-$ & $\bullet$ & $\bullet$   \\
 D8: &  $-$ & $-$ & $-$ & $-$ &  $\bullet$ & $-$ & $-$  \\
\end{tabular}
\caption{Brane configuration of the Sakai-Sugimoto model}
\end{center}

%

The specific embedding of the D8-branes in the bulk depends on the asymptotic separation of the stacks (and also on any distribution of matter on the branes), but we will focus exclusively on the case where the two asymptotic parts of the D8-brane stack sit at opposite sides of the D8 circle, in which case each side simply extends to the tip of the cigar along a line of constant $x_4$. Unlike the adjoint fields, the fundamental fields, which arise from 4-8 strings, do not start out supersymmetric, and include massless fermions along with heavy bosons.

Let us compare this case to a different theory, where supersymmetry is not broken in the flavour sector. To describe the embedding of the probe brane in that case, let us first introduce  new coordinates which are more adapted to supersymmetry. To this end, introduce a coordinate $\rho$ which makes second term in Eqn. (\ref{D4metric}) manifestly conformally Euclidean: We choose a reparameterization $u(\rho)$ that solves
\begin{equation}
\frac{g_\Omega}{g_{\rho\rho}} = \frac{u^2 f(U)}{\left( \frac{dU}{d\rho}\right)^2} = \rho^2,
\end{equation}
i.e.,
\begin{equation}
U(\rho) =
\left( \rho^\frac{3}{2} + \frac{U_0^3}{4\rho^\frac{3}{2}} \right)^\frac{2}{3} =
\rho \left( 1 + \frac{U_0^3}{4\rho^3} \right)^\frac{2}{3},
\end{equation}
which leads to the form of the metric in \cite{Kruczenski2004}. The background metric (Eqn. \ref{D4metric}) can now be written entirely in terms of \(\rho\), as
\begin{equation}
ds^2 = \left(\frac{\rho}{R_4} \right)^\frac{3}{2}\left( f_+(\rho) \eta_{\mu\nu} dx^\mu dx^\nu + \frac{f_-^2(\rho)}{f_+(\rho)} dx_4^2\right) + \frac{ R_4^\frac{3}{2} U(\rho)^\frac{1}{2} }{ \rho^2 } \left(d\rho^2 + \rho^2 d\Omega_4^2 \right),
\end{equation}
where we have defined the following functions for convenience:
\begin{equation}
    f_\pm(\rho) = \left(1\pm\frac{U_0^3}{4\rho^3} \right) = \left(1\pm\frac{\rho_0^3}{\rho^3} \right).
\end{equation}
Finally, let us refactor the conformally Euclidean ${\mathbb R}^5$ into $\mathbb R^1 \times\mathbb  R^4$, letting $\rho^2 = \lambda^2 + r^2$, so that
\begin{equation}
d\rho^2 + \rho^2 d\Omega_4^2 = dr^2+\left(d\lambda^2 + \lambda^2 d\Omega_3^2\right).
\end{equation}
In these coordinates, the embedding of the probe branes in the geometry is described by the following table:
\begin{center}
\begin{tabular}{c|cccccccc}
     & $x^0$ & $x^1$ & $x^2$ & $x^3$ & $(x^4)$ & $r$ & $\lambda$ & $\Omega_3$ \\
     \hline
 D4: & $-$ & $-$ & $-$ & $-$ & $-$ & $\bullet$ & $\bullet$  & $\bullet$  \\
 D8: &  $-$ & $-$ & $-$ & $-$ & $-$ & $\bullet$ & $-$ & $-$  \\
\end{tabular}
\caption{Brane configuration with boson-fermion degeneracy}
\end{center}
%
%

Compared to the Sakai-Sugimoto model, the adjoint sector is unaltered, while the fundamental sector now preserves $\mathcal{N}=1$ supersymmetry (before compactification). Both fundamental scalars and fermions now acquire mass of order $M_{KK}$. The theory is pure YM at sufficiently low energies, and becomes approximately supersymmetric at the KK scale. This is a natural theory to compare to the Sakai-Sugimoto model: on the one hand the gravity duals are fairly similar to each other, but on the other hand the field theories, especially at finite chemical potential, are expected to be radically different due to the difference in both the massless and the light (QCD scale) matter content.

\subsection{Baryons}
Configurations with non-zero baryon charge density require sources for $A_0$, the gauge field on the D8-branes. The basic source for $A_0$ is the endpoint of a fundamental string. In order to have some net charge,  we need a source for fundamental strings in the bulk. In our background we can get a density of charge on the D8-brane by having a density of D4-branes wrapped on $S^4$ in the bulk, with $N_c$ strings stretching between each D4-brane and the D8-brane.
In the case where we have $N_f>1$ D8-branes, our baryon branes can dissolve in the D8-branes  and show up as instantons. The question of which of these two pictures is more appropriate is a dynamical one.

In order to work at finite chemical potential $\mu$ for baryon number, we need to turn on non-normalizable mode for the operator dual to the baryon number. We require then that the value of the gauge potential $A_0$ approaches the constant $\mu$ in both asymptotic regions of the D8-brane.

\section{One Flavour Physics}

In this section we study the physics of both  models at finite chemical potential, in the simpler case of a single quark flavour. Since we have only a single D8-brane in the bulk, we can use the Abelian Born-Infeld action for our analysis. Baryons are represented by pointlike charges on the D8-branes that source the electrostatic potential; configurations with smooth  baryon density correspond to having some density of these pointlike instantons on the D8-brane.

The dynamics of the probe D8-branes are governed by the  action:
\[
S_{DBI} = -\mu_8 \int d^9 \sigma e^{-\phi} \sqrt{-\det(g_{ab} +  2 \pi \alpha'F_{ab})}
\]
Here \(g_{ab}\) is the induced metric on the brane, and $F_{ab}$ is the field strength on the brane's worldvolume. There is also a Wess-Zumino term, which will be irrelevant for the discussion.
\subsection{ The Sakai-Sugimoto Model}
Let us discuss the Sakai-Sugimoto model first. To simplify the  action, we can work in the ``static gauge'', by identifying the worldvolume and spacetime coordinates in the sphere and the field theory directions, as well as in the radial direction $U$. Additionally we choose to concentrate on the case where $x_4$ is constant. We are interested only in time-independent configurations which are homogeneous and isotropic in the spatial directions of the field theory (labeled by indices $i,j,k$). For details of the derivation see \cite{Rozalietal2008}.

For any value of the chemical potential, which is the asymptotic value of the gauge potential $A_0$, the constant configuration $A_0(\lambda)=\mu$ is always a solution of the equations of motion. For this solution there is no electric flux, and by Gauss law we also have zero baryon density. Whether or not other solutions exist, with some non-zero baryon density, depends on the value of $\mu$. This leads to the existence of distinct phases as function of $\mu$ and phase transitions between them.

Once we have some non-zero baryon density, in our setup the embedding of the branes does not change. However, to find the configuration of the gauge field $A_0$ it is important to include the energy contributions from the electric field on the branes. Starting with the DBI action above, we can derive the 3+1 dimensional energy density via a Legendre transformation.  We find
\be
\label{fluxenergy}
{\cal E}_{flux} = \frac{\mu_8 }{ g_s} \frac{8 }{ 3} \pi^2 R_4^ \frac{3 }{ 2} \int dU  \frac{  U^\frac{5 }{ 2}  }{ \sqrt{f \left(1-2 \pi \alpha'f(U)(\partial_U {A_0})^2 \right)}}   \; .
\ee

As a first approximation, we make the simplifying assumption that all the pointlike instantons sit at $U=U_0$.
In this simple approximation, the energy from the electric flux $E = \frac{\delta S}{\delta (\partial_U A)}$ (from both halves of the D8-brane) is
\bea
{\cal E}_{flux} &=& 2 \cdot \frac{\mu_8 }{ g_s} \frac{8 }{ 3} \pi^2  R_4^\frac{3 }{ 2} \int_{U_0}^\infty dU \frac{ U^\frac{5 }{ 2} }{ \sqrt{f}} \left( \sqrt{1 + \frac{E^2 }{ U^5 }} - 1 \right), \cr
&=&  \frac{\mu_8 }{ g_s} \frac{16}{ 3} \pi^2   R_4^\frac{3 }{ 2} U_0^\frac{7 }{ 2} h (e),
\eea
where we have subtracted off the $E=0$ contribution from the brane tension, and  defined $e = E/U_0^\frac{5 }{ 2}$ and
\[
h(e) = \int_1^\infty dx (\sqrt{x^5 + e^2} - x^\frac{5 }{ 2}) \frac{1 }{ \sqrt{1-1/x^3}} \; .
\]
Meanwhile, the energy from the charges in the electrostatic potential and the masses of the pointlike instantons combine to give a term
\[
{\cal E}_{charge} = -(\mu - \mu_c) n_B \; .
\] where $\mu_c$ is a positive constant.

Defining $
\tilde{\mu} = \frac {6 \pi \alpha' \mu }{ U_0} $ and using the relation between $n_B$ and $E$ (which is the DBI version of Gauss law, relating the charge to the electric flux), the total energy may be written as
\[
{\cal E} = \frac{\mu_8 }{ g_s} \frac{16 }{ 3} \pi^2 R_4^\frac{3 }{ 2} U_0^\frac{7 }{ 2}\left(h(e) - \frac{1 }{ 3}(\tilde{\mu} - 1) e \right).
\]
From this, we find that there is a non-trivial solution provided  $\mu>\mu_c$, the total  energy is then minimized when
\[
\frac{1 }{ 3} (\tilde{\mu} - 1) = h'(e) \; .
\]
This can be inverted to determine the relationship between $n_B$  and $\mu$ above the transition, we find that the chemical potential exhibits the following asymptotic behavior:
\begin{align}
&n_B \propto \mu - \mu_c &  {\rm small \;} \mu - \mu_c,\\
&n_B \propto \mu^\frac{5 }{ 2} & {\rm large \;} \mu - \mu_c.
\end{align}

In summary, for a given value of the chemical potential greater than the critical value, we find some preferred distribution of charges on the D8-brane. The total baryon density for a given value of $\mu$ may be read off from the asymptotic value of the electric flux, and the result increases smoothly from $0$ above the critical chemical potential, approaching an asymptotic behavior $n_B \propto \mu^\frac{5 }{ 2}$.
With a more careful analysis, allowing for dynamical charge distribution, it was  found  \cite{Rozalietal2008} that the qualitative behavior of $n_B(\mu)$ is the same as in the simplified model of the previous section, though the numerical coefficients come out different.

Additionally, the behavior of the energy ${\cal E}(\mu)$  can be found
\begin{align}
&{\cal E} \propto (\mu - \mu_c)^4 &  {\rm small \;} \mu - \mu_c,\\
&{\cal E} \propto \mu^\frac{7 }{ 2} & {\rm large \;} \mu - \mu_c.
\end{align}
We see that the energy vanishes at the critical point, signaling a second order phase transition. For the single flavour case, even allowing for dynamical charge distribution,
the transition to nuclear matter is continuous, unlike QCD, but it may be expected that the single flavour case is different due to the absence of pions which usually play a crucial role in interactions between nucleons. Indeed, the details found in \cite{Rozalietal2008} are different for the multi-flavour case, the transition is first order, and the critical exponents at large $\mu$ are also different
\begin{align}
&{n_B} \propto \mu^3 \\
&{\cal E} \propto \mu^4
\end{align}
Note that these are the powers appropriate for free fermions, and we would like to understand this point better.

\subsection{Supersymmetric Matter Content}

Let us now analogously develop the second model, in which there is a Bose-Fermi degeneracy in the fundamental matter. For the worldvolume coordinates of the D8-brane we choose the ``static gauge'' except for \(r(\lambda)\). Thus, the worldvolume coordinates are \(    \{x^0...x^4, \lambda, \Omega_3\}\).  The induced metric is on the D8-brane is
\begin{equation}
ds_8^2 = \left(\frac{\rho}{R_4} \right)^\frac{3}{2}\left( f_+(\rho) \eta_{\mu\nu} dx^\mu dx^\nu + \frac{f_-^2(\rho)}{f_+(\rho)} dx_4^2\right) + \frac{ R_4^\frac{3}{2} U^\frac{1}{2} }{ \rho^2 } \left[ (1 + \partial_\lambda r^2) d\lambda^2 + \lambda^2 d\Omega_3^2 \right].
\end{equation}
Let us allow one component of the gauge potential on the D8 world-volume to be non-zero, in response to a source of \(U(1)\) charge that is concentrated at a point \(U = U_0\). Let \(A_0 = A_0(\lambda)\) so that only \(F_{\lambda 0} = \partial_\lambda A_0\) is non-zero. We need to evaluate the following determinant:
\begin{align}
\det (g + F) 
&= - f_+(\rho)^2 f_-(\rho)^2 \frac{u^\frac{3}{2} \lambda^6}{R^\frac{3}{2}} \left[-\partial_\lambda A_0(\lambda)^2+ f_+(\rho)^{\frac{4}{3}} (1+\partial_\lambda r(\lambda)^2) \right](d\Omega_3)^2.
\end{align}
The DBI action governing the D8 world-volume is therefore given by
\begin{equation}
S_8 = -\frac{2\pi^2 \mu_8 V_5}{g_s} \int\!d\lambda\, f_+(\rho) f_-(\rho) \lambda^3 \left[-\partial_\lambda A_0(\lambda)^2+ f_+(\rho)^{\frac{4}{3}} (1+\partial_\lambda r(\lambda)^2) \right]^\frac{1}{2}.
\end{equation}

This system does not admit a trivial (constant) solution in the coordinates that we have chosen, and so we will require a careful analysis of the boundary conditions due to the addition of baryons in order to determine the most favourable embedding. Let the number density \(\rho_B(\lambda) = n_B \delta(\lambda)\) represent a localized source of baryons. We minimally couple baryons to the gauge field via addition of the term
\begin{align}
\mathcal{S}_{charge} &= N_c \int\!d\lambda\, A_0(\lambda) \rho_B(\lambda),\\
	&= N_c n_B \mu - N_c n_B\int_0^\infty\!d\lambda\,\partial_\lambda A_0(\lambda).
\end{align}
%
The kinetic contribution of the baryons is~\cite{Rozalietal2008}
\begin{align}
    S_{mass} 
 &= -\frac{\mu_4}{g_s}\frac{8\pi}{3}\pi^2 R_4^3 n_B \int\!d\lambda\, U(\lambda) \delta(\lambda) = -n_B N_c M_B.
 \end{align}

To find the equations of motion, we note that the variable \(A\) is effectively cyclic, except at the boundary, so that by varying \(A\), we find again that the flux
\begin{equation}
E = \frac{\delta S}{\delta \partial_\lambda A_0} = \frac{2\pi^2 \mu_8}{g_s} \frac{ f_+(\rho) f_-(\rho) \lambda^3 }{ \sqrt{-\partial_\lambda A_0(\lambda)^2+ f_+(\rho)^{\frac{4}{3}} (1+\partial_\lambda r(\lambda)^2) } } \partial_\lambda A_0(\lambda),
\end{equation}
is conserved (more precisely, \(E = N_c n_B\)). Thus we have the equation of motion
\begin{equation}
\partial_\lambda A_0(\lambda) = E \sqrt{ \frac{f_+(\rho)^\frac{4}{3} (1+\partial_\lambda r(\lambda)^2) }{ \frac{4\pi^4 \mu_8^2}{g_s^2} f_+(\rho)^2 f_-(\rho)^2 \lambda^6 + E^2 } }.
\end{equation}
Roughly this corresponds to the following asymptotics as \(\lambda \to \infty\):
\begin{equation}
A_0(\lambda) \to \mu - \frac{g_s}{2\pi^2\mu_8} \frac{E}{2\lambda^2},
\end{equation}
where \(E = n_B N_c\)~\cite{Kobayashi:2006sb}. To simplify calculations, we perform a Legendre transform with respect to \( \partial_\lambda A_0(\lambda)\): Define
\begin{align}
S^*_8	&= S_8- \int\!d\lambda E \partial_\lambda A_0(E),\nonumber\\
	&= - \frac{2\pi^2 \mu_8}{g_s} \int\!d\lambda\,f_+(\rho)^\frac{2}{3} \sqrt{ 1+\partial_\lambda r(\lambda)^2 } \sqrt{f_+(\rho)^2 f_-(\rho)^2 \lambda^6 + \tilde E^2}.
\end{align}

where we have defined
\begin{equation}
\tilde{E} = \frac{g_s}{2\pi^2 \mu_8} E = \frac{E}{k_8} = \frac{N_c n_B}{k_8}.
\end{equation}
	
The equations of motion for \(r(\lambda)\) follow from the action
\begin{align}
S^* &= S_8^* + S_{mass},\nonumber\\
&= S_8^* -\frac{\mu_4}{g_s}\frac{8\pi}{3}\pi^2 R_4^3 n_B \int\!d\lambda\, U(\lambda) \delta(\lambda)
\end{align}
To obtain  the boundary condition on \(\partial_\lambda r(\lambda = 0)\), we formally extend \(\lambda\) to be negative and extremise the ``doubled'' system. Then
\begin{equation}
 \frac{\partial}{\partial \lambda} \left( \frac{\delta S^*}{\delta \,\partial_\lambda r(\lambda) }\right) - \frac{\delta S^*}{\delta r(\lambda)} = 0.
\end{equation}
We may integrate from \(-\epsilon\) to \(+\epsilon\) and find
\begin{equation}
    2 \frac{\delta S^*_8}{\delta \partial_\lambda r (\epsilon) } - O(\epsilon) + \frac{\mu_4}{g_s}\frac{8\pi}{3}\pi^2 R_4^3 n_B \int_{-\epsilon}^{+\epsilon}\!d\lambda\, \frac{\partial U}{\partial r} (\lambda) \delta(\lambda) = 0,
\end{equation}
and, taking the limit \(\epsilon \to 0\), we obtain the required boundary condition:
\begin{align}
\label{bdrycondition}
    2 \frac{\delta S^*_8}{\delta \, \partial_\lambda r(0)} 
 &= - \frac{1}{6\pi \alpha'} k_8 \tilde{E} \tfrac{\partial U}{\partial r} (0),
\end{align}
where \(k_8 = \frac{2\pi^2 \mu_8}{g_s} (2\pi R)\).
We may use this to solve for \(\partial_\lambda r(0)\) in terms of \(r(0)\) and \(n_B\). Here it is important to note that \( \frac{\partial U}{\partial r} \) is strictly negative, so that the solution consistent with Eqn. (\ref{bdrycondition}) must be selected. With the boundary condition determined, we may ignore the baryon and solve the homogeneous equations of motion for \(r(\lambda)\) numerically. After specifying \(r(\infty)\) as the bare quark mass, we can then solve for \(r(0)\) in terms of \(r(\infty)\). Each embedding is therefore uniquely specified by the additional parameters \(\mu\) and \(E\), or, equivalently, \(\mu\) and \(n_B\).

\begin{figure*}[htp]
\label{mu_nb_fig_2}
\centering
\includegraphics[scale=0.6]{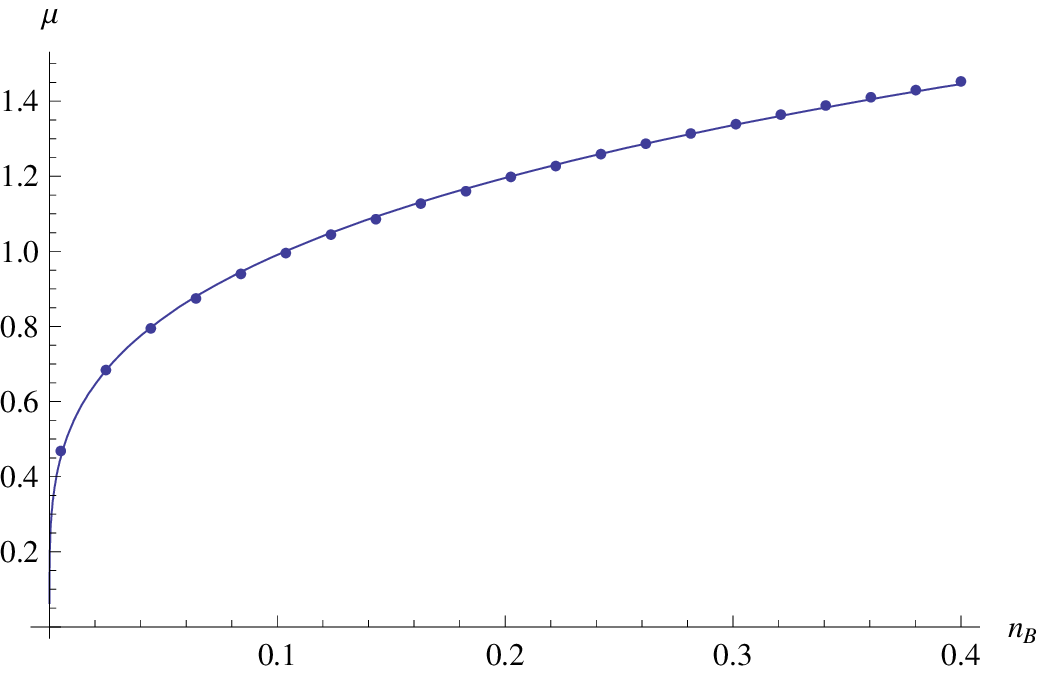}
\caption[Power-law dependence of chemical potential on baryon number]{Power-law dependence of chemical potential on baryon number with best fit indicated by solid line. Shown \(r_\infty = \frac{7}{5} \rho_0\).}
\end{figure*}

The relationship between \(n_B\) and \(\mu\) is fixed by requiring that the total energy of the configuration be minimized. Since the configuration is static, the energy is equal to minus the Lagrangian evaluated over a solution to the equations of motion, i.e.,
\begin{equation}
\mathcal{E} = - \frac{1}{V_4} \left( S_8 + S_{charge} + S_{mass} \right),
\end{equation}
where \(V_4 = \int\!dx^0 \int\!d^3x\). For numerical purposes, we work in string units, noting that the number of integrations eliminate all powers of \(\ell_s\). The first two terms partly combine to form
\begin{equation}
\mathcal{E}_{flux} = k_8 \int\!d\lambda\,f_+(\rho)^\frac{2}{3} \sqrt{ 1+\partial_\lambda r(\lambda)^2 } \sqrt{f_+(\rho)^2 f_-(\rho)^2 \lambda^6 + \tilde E^2},
\end{equation}
while the remainder is given by \( -\mu N_c n_B \) and \(\mathcal{E}_{mass} = \frac{N_c}{6\pi} U_{min} n_B = n_B N_c M_B\). All contributions together give the energy in terms of the convenient variable \(\tilde{E}\), which is then to be minimized over the family of solutions \( r(\lambda)\) parameterized by this variable.
\begin{equation}
\frac{\mathcal{E}}{k_8} = - \left(\mu - M_B \right) \tilde{E} + \int_0^\infty \!d\lambda\,f_+(\rho)^\frac{2}{3} \sqrt{ 1+\partial_\lambda r(\lambda)^2 } \sqrt{f_+(\rho)^2 f_-(\rho)^2 \lambda^6 + \tilde E^2}.
\end{equation}
For a fixed value of \(\mu\), we may seek the value of \(\tilde{E}\) which solves
\begin{equation}
\label{mufromnb_2}
\mu = \frac{d}{d\tilde{E}} \left( M_B\tilde{E} + \int_0^\infty \!d\lambda\,f_+(\rho)^\frac{2}{3} \sqrt{ 1+\partial_\lambda r(\lambda)^2 } \sqrt{f_+(\rho)^2 f_-(\rho)^2 \lambda^6 + \tilde E^2} \right).
\end{equation}
As it stands, Eqn. gives the chemical potential in terms of baryon number, without modification. All quantities may be numerically evaluated, and the behaviour of \( \mu(n_B)\) for small values of \(n_B\) is shown in Fig. (\ref{mu_nb_fig_2}), along with a best fit to $\mu -\mu_c \sim n_B^{\frac{2}{7}}$, with \( \mu_c \approx 0.32\, M_B(E=0)\).

The calculation presented so far has assumed a non-zero flux \(E\). There is always another solution, with \(E=0\), where \(\mu\) may take on any value, corresponding to a constant gauge potential \(A_0(\lambda) = \mu\) in the absence of sources. For values \(\mu < \mu_c\), this solution is energetically favourable, after which baryons begin to accumulate. This situation has been analyzed using the best-fit data obtained above, which indicates a smooth transition of second or higher order into the nuclear matter phase.

It is interesting to compare the behavior here to that found in the Sakai-Sugimoto model. We have found that the details of the transition to nuclear matter are identical; it is smooth in both cases, with same critical exponent near the transition point. This is somewhat surprising given that the massless matter content is very different in both cases. Perhaps the \(N_f = 1\) case cannot be expected to be very different from pure YM, and quanlitative differences only arise when massless pions become part of the story.

On the other hand, we do find different critical exponents at large chemical potential. This is expected, as in our model there is an approximate fermion-boson degeneracy at high enough energy, which is sufficient to cause the model to behave differently in the regime of high chemical potential.

Most interestingly, we find that the binding energy per nucleon is of the order of magnitude it is expected to be, the baryon mass itself. In other words the baryons form a deep bound state, unlike the case of the Sakai-Sugimoto model, and of real-world QCD. It is interesting to understand further the mechanism for forming shallow bound states, and the dependence of the binding energy on specifics of the model.

\section{Other Results for the Sakai-Sugimoto Model}

We review here other results obtained in \cite{Rozalietal2008}, it would be interesting to compare those to other models, to probe  their dependence on the precise matter content and other details of the theory.

\subsection{Transition to Nuclear Matter}

We have seen that turning on a chemical potential in the gauge theory corresponds to including boundary conditions $A_0 = \mu$ for the two asymptotic regions of the D8-brane. For any $\mu$, one solution consistent with these boundary conditions is to have constant $A_0$ everywhere on the brane.  This represents the vacuum configuration in the field theory. However, beyond a certain critical chemical potential, this solution is unstable to the condensation of baryons. This is manifested by the existence of another classical solution with non-trivial flux, which dominates the thermodynamics beyond the critical value $\mu_c$. A first approximation to this critical value is the baryon mass, since it is at this point where it becomes energetically favorable to add single baryons to the vacuum. In fact, the critical value is somewhat lower, since the baryons have  binding energy. Therefore the critical value $\mu_c$ can be used to estimate the binding energy per baryon.

The binding energy per nucleon is estimated in real QCD to be
\be
\label{bvol}
b_{vol} = 16 \; {\rm MeV} \; .
\ee
which is notably small compared to the QCD scale.

In the Sakai-Sugimoto model, for a single flavour the transition is second order, thus involving no latent energy. Within the localized source approximation the critical point is precisely at the baryon mass $\mu_c=M_B$, therefore in that case the binding energy per nucleon vanishes. However, in the case with $N_f > 1$, we can have nonsingular instantons on the $N_f$ coincident D8-branes, and the minimum energy configurations for large enough $\mu$ are should involve smooth configurations of the non- Abelian gauge field carrying an instanton density. This is the mechanism by which the baryons bind to each other in the holographic context.

For many flavours, the Sakai-Sugimoto model displays a first order transition to nuclear matter at some calculable critical chemical potential $\mu_c$. We find the binding energy per nucleon
\[
E_{bind} = M_B - \mu_c  \approx \frac{N_c }{ 27 \pi} M_{KK} (c' - c) \; ,
\]
where $c,c'$ are unknown constants of order one, and the result is  insensitive to the value of $\lambda$ for large $\lambda$. Since we also know that this binding energy approaches some constant value in the limit of small $\lambda$, then assuming a smooth behavior at intermediate values of $\lambda$, we can treat the large $\lambda$ result as a prediction for the order of magnitude of the QCD result. Noting that $M_{KK} \approx \Lambda_{QCD}$ for large $\lambda$, the value of the binding energy per nucleon extrapolated to $N_c=3$ becomes
\[
E_{bind} = \frac{1 }{ 9 \pi} \Lambda_{QCD} (c'-c) \approx 7 \; {\rm MeV} (c'-c)
\]
thus we obtain the same order of magnitude as the $QCD$ result. In particular the binding energy is anomalously small compared to the QCD scale. It would be interesting to understand the mechanism for this, and how sensitive is the result to exact details of the model.

\subsection {Chiral Density Waves}

One significant difference between the large-$N_c$ theory and ordinary QCD is the expected behavior at asymptotically large values of the chemical potential \cite{Deryagin:1992rw}. In both cases, we have attractive interactions between excitations on the Fermi surface that result in an instability, but the nature of the resulting condensates is different. Whereas for $N_c=3$ the instability is a BCS-type instability, believed to lead to a color superconductor phase, the dominant instability at large-$N_c$ is toward the formation of ``chiral density waves'', inhomogeneous perturbations in the chiral condensate with wave number of order twice the chemical potential. This suggests that the ground state for large-$N_c$ QCD at large enough chemical potential is inhomogeneous, however the nature of the true ground state remains mysterious.

In the holographic model of Sakai and Sugimoto, it was found in \cite{Rozalietal2008} that there are no allowed configurations of the D8-brane gauge field that are spatially homogeneous in the three field theory directions such that the net energy density and baryon density in the field theory are both finite. Thus, any phase with finite baryon density is necessarily spatially inhomogeneous. This has a simple interpretation: it suggests that at large-$N_c$, the nucleons retain their individual identities for any value of the chemical potential.  This suggests that the chiral density wave instability of the quark Fermi surface in large-$N_c$ QCD simply indicates that the quarks want to bind into nucleons even at asymptotically large densities. It is interesting to understand this phenomena further.

\subsection{Holographic Fermi Surface}

It is interesting to observe some characteristics of the charge distribution, once we allow it to become dynamical. This is summarized by a density of pointlike instantons, in the single flavour case, and by the distribution of the instanton density in the non-Abelian case. In all those cases,  the charge distribution has a sharp edge in the holographic radial direction, at $U=U_{max}$ which progresses further and further towards the UV  as the chemical potential is increased. In the field theory picture, the radial direction represents an energy scale, so the charge distribution we find in the bulk should be related to the spectrum of energies for the condensed baryons. The edge of the distribution may be a bulk manifestation of a (quark) Fermi surface. It would be interesting to check this interpretation by comparing to other models with flavoured massless bosons, where such Fermi surface is not expected.

\section*{Acknowledgements}
We thank Brian Shieh, Mark van Raamsdonk and Jackson Wu for valuable and enjoyable discussions, and collaboration on \cite{Rozalietal2008}. The work is supported in part by NSERC discovery grant and by the National Science and Engineering Research Council of Canada.

\bibliographystyle{unsrt}
\bibliography{paper}        
\end{document}